# Partial vs. integer electron transfer in molecular assemblies: On the importance of a multideterminant theoretical description and the necessity to find a solution within DFT

Victor Geskin[1], Robert Stadler[2], Jérôme Cornil[1]

[1]Laboratory for Chemistry of Novel Materials, University of Mons, Place du Parc 20, B-7000 Mons, Belgium
[2]Department of Physical Chemistry, University of Vienna, Sensengasse 8/7, A-1090 Vienna, Austria

**Abstract.** Nonequilibrium Green's function techniques (NEGF) combined with density functional theory (DFT) calculations have become a standard tool for the description of electron transport through single molecule nanojunctions in the coherent tunneling (CT) regime. However, the applicability of these methods for transport in the Coulomb blockade (CB) regime is questionable. For a molecular assembly model, with multideterminant calculations as a benchmark, we show how a closed-shell ansatz, the usual ingredient of mean-field methods, fails to properly describe the step like electron-transfer characteristic in weakly coupled systems. Detailed analysis of this misbehavior allows us to propose a practical scheme to extract the addition energies in the CB regime for single-molecule junctions from NEGF DFT within the local-density approximation (closed shell). We show also that electrostatic screening effects are taken into account within this simple approach.

## INTRODUCTION

Interest in electron transfer between nanoscale contacts is intense due to the advent of the technologically motivated field of molecular electronics and recent progress in experimental techniques for manipulating and contacting individual molecules. Electron transport at this scale can operate between two limiting regimes, namely, coherent transport (CT) and Coulomb blockade (CB) that can be considered as complementary in a quantum mechanical sense: the carrier is delocalized throughout the electrode-molecule-electrode system in CT but localized at one of these parts in CB, which implies integer electron transfer.

Nonequilibrium Green's-function formalism (NEGF) methods applied to the theoretical description of electron transport through single-molecule junctions are typically implemented in combination with density-functional theory (DFT) and a closed-shell ansatz (double occupation of the orbitals). Well suited for CT, it is hardly applicable to CB. Indeed, a complete electron transfer between two initially closed-shell moieties leads to an open-shell singlet state. However, the behavior, or rather misbehavior, of a closed-shell solution merits analysis as a usual ingredient of mean-field methods in electron-transport calculations, and this is the first goal of our work. We show numerically and derive analytically using molecular models [1, 2] how a closed-shell ansatz always predicts partial electron transfer, which can lead to qualitatively erroneous assessments. The partial and integer electron transfer (equivalent to CT and CB situations) are automatically discerned, and the spin purity is assured, by application of a multideterminant wave function ansatz that serves as a benchmark. Our second goal is to explore whether correct understanding of the closed-shell DFT behavior can help in extracting useful information from these calculations even in CB regime for molecular assemblies and molecular junctions. We devise a practical scheme to estimate addition energies from standard NEGF-DFT calculations on metal-molecule-metal open systems [3] and demonstrate that screening effects renormalizing molecular levels are reasonably taken into account within it [4].

## MOLECULAR MODEL FOR CHARGE TRANSFER IN A JUNCTION

Let us replace infinite metal electrodes of an electrode-molecule-electrode system by closed-shell molecules and focus on a molecular triad as a model for electron-transfer processes triggered by an external electric field [1, 2]. This approach does not allow for the definition of a conductance or current and relates rather to e.g. spectroscopic capacitance experiments, which can equally demonstrate gate-induced charge-quantized electron transfer. However, its simplicity allows us to apply all the rigor and machinery of quantum chemistry to the problem.

For a weakly coupled molecular triad, a series of full-electron-transfer steps is found in the multideterminant configuration interaction (CI, Figure 1a) description, as expected. On the other hand, the electron transfer is continuous, and only fractional charges are found on the moieties within closed-shell restricted Hartree-Fock (RHF (Figure 1b) up to the point where two electrons have been exchanged. If these results were taken from experimental I-V curves instead of theoretical calculations, one would interpret one-electron steps (as found with CI) as an indicator for the CB regime and would associate the monotonic evolution (as obtained from RHF) with CT. We thus reach the conclusion that the indiscriminate application of the closed-shell ansatz may lead to a physically wrong assignment of the transport regime.

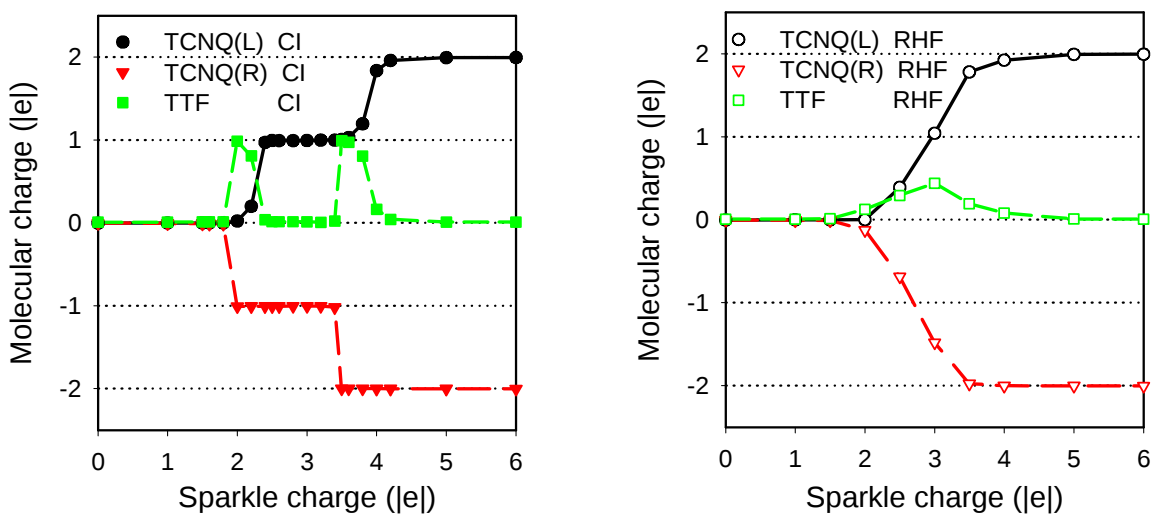

**FIGURE 1.** Intermolecular charge transfer in a cofacial TCNQ-TTF-TCNQ triad induced by external electric field of "sparkle" charges of opposite sign as obtained from multideterminant CASCI (a) and RHF calculations with the AM1 Hamiltonian.

Note, however, that the very onset of continuous charge transfer within RHF coincides with the field strength required to induce a full one-electron transfer within CI. This deep correspondence suggests that an RHF result might probably be useful even beyond the domain of the strict applicability of the method, if properly assessed, and motivates a deeper analysis by means of an analytical model. The basic features of the RHF solution behavior are captured already within a model for charge transfer under Zero differential overlap between 2 MOs (one on the donor, one on the acceptor) bearing a total of 2 electrons (Z22) [1]. This model explains (i) the sharp onset of continuous charge transfer in RHF as a function of the external field, (ii) the reason for the unphysical linearity of the field dependence of the charges on the donor and the acceptor units in RHF, (iii) the origin of the HOMO-LUMO gap in the RHF mixing region, (iv) why the RHF charge-transfer onset coincides with the full one-electron transfer in CI, and finally (v) how CI corrects for the described artifacts of RHF.

It is instructive to consider the role of the HOMO-LUMO gap in electron-transfer processes and its meaning within HF and DFT ideologies. The RHF calculations show that charge transfer in weakly coupled molecular assemblies onsets when this gap decreases to a certain threshold and then remains constant until two electrons are transferred (Figure 2). Our Z22 model allows us to relate the RHF critical gap $\Delta_{CT\,RHF}$ to the effective electron-transfer distance $d_{eff}$ as $d_{eff} = 1/\Delta_{CT\,RHF}$ (where atomic units have been assumed and related conversion factors have been omitted). The values thus obtained reasonably characterize the molecular systems studied. Note that this gap has nothing to do with avoided level crossing. Within DFT, charge transfer requires a zero HOMO-LUMO gap, and a virtually linear relation between the threshold MO gap and the percentage of HF exchange is characteristic for hybrid DFT.

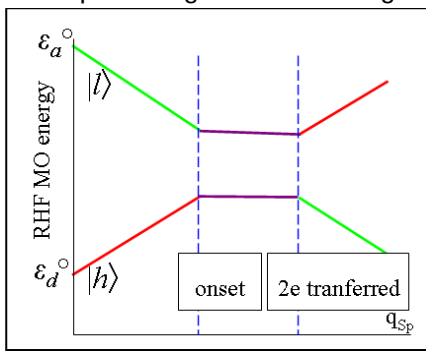

**FIGURE 2.** Schematic RHF frontier MO energy evolution during charge transfer in a weakly coupled molecular dyad

Wave-function theory provides a general solution for charge-transfer situations, namely, with its multideterminant methods. For practical reasons, we have explored [1] the open-shell (spin-polarized) solutions of DFT with hybrid functionals and found that qualitatively correct step like behavior of charging curves can be obtained, although the quantitative values for the onsets still differ from the multideterminant benchmarks. There is a complication in obtaining these solutions: the spin symmetry of the initial guess, closed-shell by default for even number of electrons in the absence of degenerate HOMOs, has to be broken in some way in order to reach the energetically lowered and correct open-shell solution for systems where such a solution is likely to exist.

There is, however, a different way to approach the problem within DFT, where it becomes possible to derive quantitatively realistic values for CB addition energies from closed-shell calculations although the charging curves are qualitatively wrong.

## APPROXIMATE THEORETICAL DESCRIPTION OF MOLECULAR JUNCTIONS IN COULOMB BLOCKADE REGIME BASED ON DFT

We performed NEGF-DFT calculations in the presence of an effective gate potential $V_{gate}$, which allows for an explicit charge transfer between one-dimensional lithium chains as electrodes, and $H_2$ and benzene molecules as the central unit [3]. The onsets for the first and second electron transfer (to or from the molecule) can be determined from the gate voltages required to move $N_{add}$ of 0.5 and 1.5 electrons (Figure 3a), respectively, which is justified by the application of the midpoint rule. The more linear the dependence of $V_{gate}$ with respect to $N_{add}$, the more precise this approximation. For the molecules above, the NEGF-DFT half-electron results compare favorably to total energy differences, experiments, and accurate GW calculations. The results obtained for these model systems show that (i) quantitatively realistic values can be derived for the addition energies $E_{add}$ associated with CB diamonds despite the fact that our calculations are carried out in the local-density approximation (LDA) and introduce a distortion in the charging process characteristic for DFT and (ii) the size of the KS gap estimated from the electronic eigenenergies of the frontier orbitals is not directly related to $E_{add}$.

Encouraged by such performance, we address more explicitly the screening by metal known to affect on the ionization and affinity level position and, therefore, addition energies of a molecule in its vicnity. We apply the midpoint rule-based approach to calculate the addition energies for a benzene molecule lying parallel to the surface of two aluminum electrodes [4] and focus in particular on the dependence of $E_{add}$ on metal-molecule distance (Figure 3b). When comparing our model setup to the experimentl situation, it is essential to keep in mind that the $E_{add}$ calculated consists of a sum of two terms: the first related to the modified molecular HOMO-LUMO gap in the junction and the second to the electrostatic capacitance of the metallic electrodes. The latter, inversely proportional to the area of the electrodes, can be safely neglected in the analysis of CB experiments on single-molecule junctions. This is not the case, however, in our calculations with their rather dense periodic lateral arrangement of molecules leading to drastic reduction of electrode surface per molecule. In order to make our results meaningful with respect to experimental values, the calculated addition energy should be corrected by substracting the geometric capacitive energy. This being done, when we compare the distance dependence of the corrected $E_{add}$ as obtained from NEGF-DFT calculations to the results provided by the image charge model, we find the deviations to be less than 20% of the total value of the former, which is quite remarkable given the approximative nature of the model.

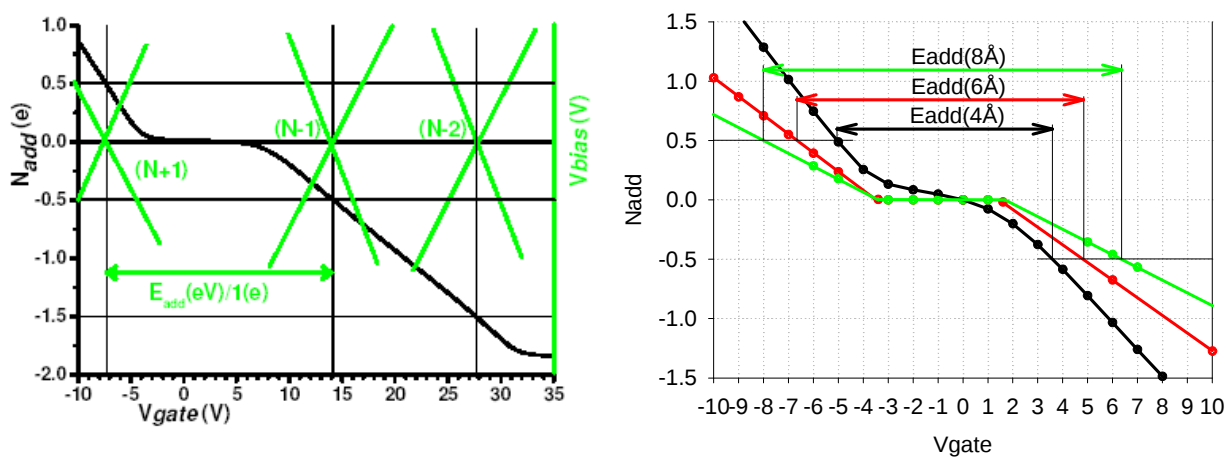

**FIGURE 3.** Evolution of the extra electrons on the $H_2$ molecule in the junction for one distance (a) and on the benzene molecule for three distances between the molecule and the electrodes (b) as a function of gate voltage. CB diamonds at the predicted threshold voltages are also schematically shown in (a).

## CONCLUSION

Theoretical NEGF DFT treatment of mesoscopic electron transport is usually limited to the closed-shell case (single-determinant spin-restricted ansatz), well-adapted to Coherent Tunneling. However, Coulomb Blockade effectively leads to an open-shell configuration. Enforcing a closed-shell solution in this case leads to qualitatively wrong results. Molecular models, for which an accurate multideterminant solution is feasible, are helpful for deep understanding of the behavior of approximate solutions in the Coulomb Blockade and Coherent Tunneling domains. Building on this understanding, a simple application of the standard NEGF DFT treatment is proposed to extract molecular Coulomb blockade addition energies for molecular junctions, taking into account the screening effects due to metal electrodes.

## ACKNOWLEDGMENTS

This research has been supported by the European Commission with the projects SINGLE (FP7/2007-2013 Grant No. 213609) and MODECOM (Grant No. NMP3-CT-2006-016434), the Interuniversity Attraction Pole Program of the Belgian Federal Science Policy Office PAI 6/27, and the Belgian National Fund for Scientific Research FNRS. R.S. is currently supported by the Austrian Science Fund FWF (Project No. P20267).